\documentclass[aps,prb,floatfix,twocolumn,superscriptaddress,showpacs,amsmath,amssymb]{revtex4}
\usepackage{graphicx}
\usepackage{epsfig} 
\usepackage{amssymb}
\usepackage{amsmath}
\usepackage{color}
\allowdisplaybreaks

\newcommand{\angstrom}{\textup{\AA}}

\begin{document}
\title{Crystal Structure Prediction of Molecular Crystals from First Principles: Are we there yet?}

\author{Cong Huy Pham}
\affiliation{Scuola Internazionale Superiore di Studi Avanzati, Via Bonomea 265, 34136 Trieste, Italy}
\affiliation{International Center for Theoretical Physics, Strada Costiera 11, 34151 Trieste, Italy}
\author{Emine Kucukbenli}
\altaffiliation{Current address: SISSA, Via Bonomea 265, 34136 Trieste, Italy}
\affiliation{\'Ecole Polytechnique F\'ed\'erale de Lausanne, CH-1015 Lausanne, Switzerland}
\author{Stefano de Gironcoli}
\email{degironc@sissa.it}
\affiliation{Scuola Internazionale Superiore di Studi Avanzati, Via Bonomea 265, 34136 Trieste, Italy}
\affiliation{CNR-IOM Democritos National Simulation Center, Via Bonomea 265, I-34136 Trieste, Italy}

\date{\today}
\begin{abstract}
Accurate molecular crystal structure prediction is a fundamental goal in academic
and industrial condensed matter research and polymorphism is arguably
the biggest obstacle on the way.
We tackle this challenge in the difficult case of the
repeatedly studied, abundantly used aminoacid Glycine that
hosts still little-known phase transitions 
and we illustrate the current state of the field through this example.
We demonstrate that the combination of recent progress in structure search algorithms
with the latest advances in the description of van der Waals interactions in Density Functional Theory,
supported by data-mining analysis, enables a leap in predictive power:
we resolve, without prior empirical input, all known phases of glycine, as well as
the structure of the previously unresolved $\zeta$ phase
after a decade of its experimental observation [Boldyreva et al. \textit{Z. Kristallogr.} \textbf{2005,} \textit{220,} 50-57].
The search for the well-established $\alpha$ phase instead reveals
the remaining challenges in exploring a polymorphic landscape.
\end{abstract}
\maketitle

\section{Introduction}
Molecular polymorphism, the observation of different crystal structures made up of the same molecules,
has been a central problem standing in the way of affordable and reliable crystal
structure prediction (CSP)
which would greatly accelerate the development of new materials
for applications in solid state chemistry, material science and pharmaceutical science \cite{Woodley2008,Desiraju2002}.
The key challenges for ab initio CSP of molecular crystals can be summarized as
\textbf{i)} the computational cost of thermodynamical exploration of a rich polymorphic phase space,
\textbf{ii)} the accuracy needed to resolve similarly-low energies among polymorphs \cite{Garnet2014},
and \textbf{iii)} the fact that kinetic factors may control the crystallization procedure
rather than thermodynamic ones \cite{Price2004}.

The past decade witnessed these challenges being tackled by the scientific community
and the progress can be followed through the blind tests organized yearly by the
Cambridge Crystallographic Data Centre \cite{CCDC,CCDC2001,CCDC2011}.
The exponential growth in the hardware performance and new,
efficient algorithms tailored for molecular crystals have allowed a wider
region of the phase space to be explored.
The increased computational performance also enabled a transition from
empirical interatomic potentials to more accurate but time consuming
quantum mechanical techniques, mainly Density Functional Theory (DFT).
This transition did not guarantee however an increase in predictive power in all cases
\cite{USPEX-molecule}:
the standard DFT functionals do not describe properly van der Waals (vdW) interactions,
which forces CSP studies to employ approximate semi-empirical corrections.
These approximations to vdW interactions strongly affect the energy ordering of explored structures,
which is a core information in predicting polymorphism.
Hence, to render CSP predictions reliable, a fully ab initio method, able to obtain an
accurate lattice energy including the vdW interactions, has been highly desirable.

Recently a breakthrough in the description of vdW interactions in DFT has been made:
many new non-local functionals that accurately describe the dispersion
interactions have been proposed and
demonstrated unprecedented success in a wide range of systems from molecules, molecular crystals
to layered materials, with a computational cost comparable to that of standard functionals \cite{vdW-DF,other-vdW}.
It has been recently shown that even in difficult cases such as glycine crystals, where
polymorphs show energy differences as little as 1 kcal/mol,
new non-local functionals can yield the correct stability ordering
as well as accurate pressure evolution \cite{vdW-glycine}.

Encouraged by these results we combine this critical progress in DFT with
recent developments in evolutionary CSP \cite{USPEX},
specifically adapted for molecular structure search \cite{USPEX-molecule}, and
perform a fully ab initio CSP search on glycine crystals,
without semi empirical corrections in the energy description,
using neither information on cell geometry nor the symmetry of the experimentally observed
polymorphs. We thus assess whether state-of-the-art ab-initio CSP can pass the challenging blind
test of exploring the phase space of polymorphic glycine.

\begin{figure}[!h]
    \centering
\includegraphics[width=0.47\textwidth]{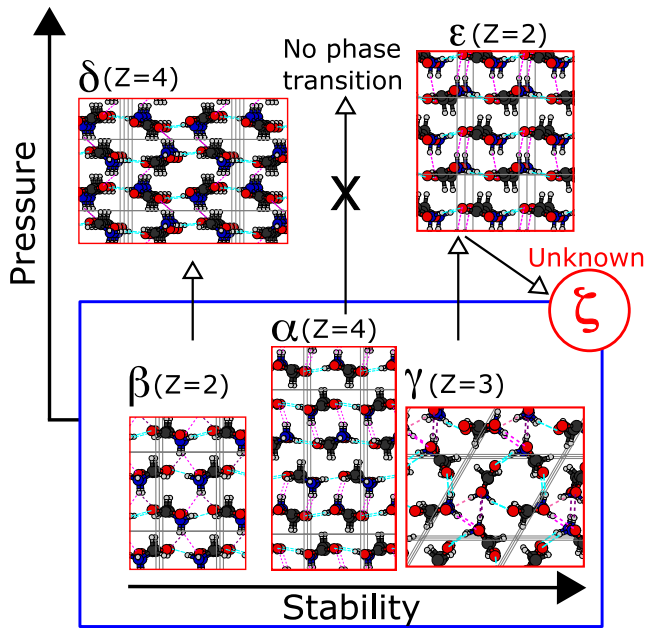}
    \caption{
(color) \textbf{Glycine polymorphism under pressure.}
The stability order of polymorphs at ambient pressure, $ \gamma > \alpha > \beta$, with indicated $Z$ molecules in unit cell, is given.
The form readily obtained by evaporation of aqueous solutions is $\alpha$-glycine,
which for long was believed to be the most stable phase instead of the later discovered ground-state phase $\gamma$.
Pressure evolution of ambient pressure phases show that
while $\gamma$ and $\beta$ phases quickly lose single crystal nature or
undergo a phase transition within a few GPa, $\alpha$ phase stays stable up to 23~GPa,
the highest pressure reached in experiments.
A reversible, hysteresis-free single-crystal to single-crystal transition
occurs from $\beta$ to $\delta$ phase at 0.76~GPa.
Single crystals of the $\gamma$ phase instead undergo an extended polymorphic transformation
in the wide range of 2.7-7.6~GPa,
to a high-pressure polymorph, the $\epsilon$ phase,
accompanied with the fragmentation of single crystals into powder.
Upon decompression, the $\epsilon$ phase is stable down to 0.62~GPa,
where a new, irreversible phase transition occurs to the
$\zeta$ phase, a new polymorph which is reported to be stable at ambient conditions
for at least three days.}
\label{fig1}
\end{figure}

Glycine, NH$_2$CH$_2$COOH, the smallest aminoacid, is an excellent test case for CSP studies
as its already rich polymorphism under ambient conditions is
amplified and becomes less understood at higher pressure (see Fig.\ref{fig1}).
A clear example to this is the $\zeta$ phase, which is reported to be stable at ambient conditions
for at least three days \cite{zeta}.  Interestingly,
despite its stability, and at least three CSP studies devoted to Glycine so far
\cite{USPEX-molecule,glycine-CSP-fail1,Chisholm2005},
a decade after its observation, the $\zeta$ phase has not been structurally resolved yet.

The complex polymorphism of glycine highlights the importance of performing
an extensive search in phase space, while practical concerns limit any
CSP study to explore primarily the lowest energy structures.
In this study we use evolutionary algorithms (EA) as implemented in the USPEX package to address
this interplay efficiently \cite{USPEX-molecule}.
We perform three test suits with Z=2, 3 or 4 glycine molecules in the crystal unit cell.
At the first generation, we start with a population of 30 random structures.
This population evolves through generations where only the thermodynamically
most stable members are allowed to 'procreate'.
The procreation operations are cross-overs of parent structures,
and mutations that involve variation of the molecular position and orientation.
The diversity of the population is guaranteed by addition of new random structures at each generation.
The highest computational cost in this workflow is due to the
ab initio geometry optimization of each structure considered.
To keep this cost well within the capacity of modern high-performance computing technologies
and within the budget of academic as well as industrial research,
we limit the evolution to 20 generations at most (see Methods for details).

\section{Methods}

\subsection{Evolutionary Search}
We use EA as implemented in the USPEX package to search for the low-energy structures
of glycine with Z=2, 3 or 4 molecules in the unit cell.
At the first generation, 30 structures are created randomly.
After energy ordering, the $20\%$ of the population that is energetically least favorable is discarded.
Among the remaining, a fingerprint analysis is performed and potential parents whose fingerprint
is within a threshold distance of 0.01 from any lower energy structure are discarded as well.
The so-determined unique structures are eligible as parents and are allowed to procreate.
The 30 new structures of the next generation are created from parents through the following operations:
heredity (cross-over of two structures) ($40\%$), softmutation (translation and rotation based on
estimate of soft vibrational modes) ($20\%$), rotation of the molecule ($20\%$), and
random structure generation ($20\%$).
In addition, the three best parents are directly cloned to the next generation.
In all simulations, the maximum number of generations was 20.

\subsection{ab initio Calculations}

For every structure generated by USPEX, the geometry and cell relaxation is performed using vdW-DF functional \cite{vdW-DF} which
was implemented in the QUANTUM ESPRESSO package \cite{QE}.
A kinetic energy cutoff of $80$ Ryd and a charge density cutoff of $560$ Ryd are used.
The Brillouin zone sampling resolution was gradually increased in three steps during relaxation: resolution of $2\pi \times 0.12~\angstrom^{-1}$,
$2\pi \times 0.10~\angstrom^{-1}$ and $2\pi \times 0.08~\angstrom^{-1}$ respectively.
Energies and geometries of the last step with the densest k-point are used throughout the study.
PAW pseudopotentials are taken from the PSLibrary project\cite{PSLibrary}.
By using this setup all structures are fully relaxed within a convergence of less than
0.1 mRy for absolute total energy,  $0.5$ mRy/a.u. for the forces on atoms and
less than $0.005$ GPa for the stress tensor.

\subsection{Cluster analysis}
The cluster analysis is performed by using single linkage clustering,
where two structures with fingerprint distance less than distance threshold $d$ are considered to belong to the same cluster.
Since USPEX definition of fingerprint does not include any information on the enthalpy of the structure,
a constraint is added such that two structures with enthalpy difference more than 0.5 kJ/mol are not allowed to form a cluster.
This constraint is found necessary only when the clustering analysis is performed for all the encountered structures,
while limiting the analysis to low enthalpy region, such constraint was not necessary as each cluster was successfully identified with
distance only.

\section{Results and Discussion}
\begin{figure}[!bh]
\centering
\includegraphics[width=0.47\textwidth]{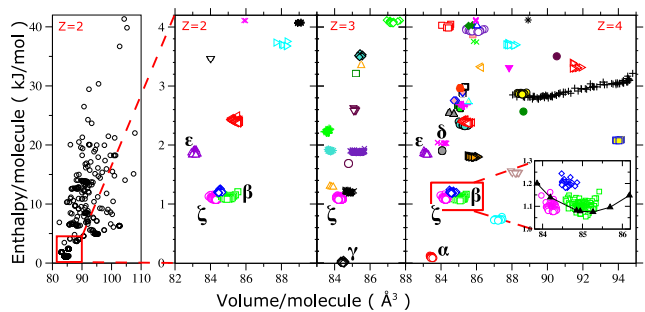}
\caption{ (color) \textbf{Results of ab initio crystal structure search for Glycine with cluster analysis.}
\textit{Left panel}: Enthalpy vs volume distribution of all encountered structures for 2 molecules per cell shows that
CSP with evolutionary algorithm allows a wide energy range to be explored while ``survival of the fittest''
algorithm keeps the focus on the thermodynamically low lying structures.
\textit{Right panels}: Expanded view of all explored structures compatible with 2, 3 and 4 molecules per cell
in the lowest 4 kJ/mol range. All known phases of Glycine are identified with the right energy ordering along
with a number of low-lying alternative polymorphs, including our prediction for the hitherto unresolved
$\zeta$ phase. As shown in the \textit{inset} of the Z=4 panel, crowding around each polymorph, when compared with
its equation of state, is compatible with numerical noise due to incomplete relaxation. The
distance-based clustering techniques adopted here are however well suited to separate and identify
the different low lying polymorphs even in presence of noise.
}
\label{fig2}
\end{figure}

The results of CSP can be visualized through the distribution of energy
as a function of volume for the structures encountered during the search.
Despite the exploration of a wide region in phase space (see left panel of Fig.\ref{fig2}),
about 40\% of all the structures lies
within 4 kJ/mol of the experimentally known ground state structure, $\gamma$.
Focusing on this region of the energy landscape as shown in the right panels of Fig.\ref{fig2},
we see structures forming islands with varying size and shapes.
This feature illustrates the added complication in the case of molecular CSP
with respect to standard inorganic solids
where a well-defined, isolated minimum would be observed for each phase.
The shape and finite size of the islands can be understood considering that
Glycine is very soft, therefore structures that are far off from the equilibrium
lattice parameters are thermodynamically penalized only slightly
as demonstrated in the inset of Fig.\ref{fig2}.
This effect, combined with the numerical noise in geometry optimization,
as well as an increased number of degrees of freedom in molecular crystals,
is enough to give rise to crowding around each polymorphic minimum.
Nevertheless islands are well separated and a clear assignment of polymorphs
can be made for most of them. This is in stark contradiction with a very recent CSP study
for glycine with empirical corrections for intermolecular interactions,
which reported that the obtained energy-volume points were not separated well enough to clearly identify each polymorph,
thus underlining the challenge of polymorphism for CSP \cite{glycine-CSP-fail1}.
In this study instead the separation between several islands are well represented down to very small
energy differences (inset of Fig.\ref{fig2}). We believe this stems from the leap in accuracy and precision reached
by the use of fully ab initio energetics together with last generation evolutionary algorithm tools.

ssibility of machine learning the polymorphs
Reliable energetics from ab initio calculations is necessary but not sufficient to guarantee a
reliable structure classification in CSP. More than one polymorph can be present
within a given extended island; or what appears to be two adjacent islands due to insufficient sampling and/or
relaxation, may actually correspond to the same packing order.
Indeed the most human-time consuming part of a CSP procedure
is known to be the stage where the output structures are comparatively
examined in order to successfully separate the essential data from the
crowd of repetitions \cite{Price2014}. Although not utilized to their full extent within CSP, concepts from data mining,
mainly clustering techniques, can be of great help in this stage of the analysis,
as we demonstrate in the following.
\begin{figure}[!h]
\centering
\includegraphics[width=0.47\textwidth]{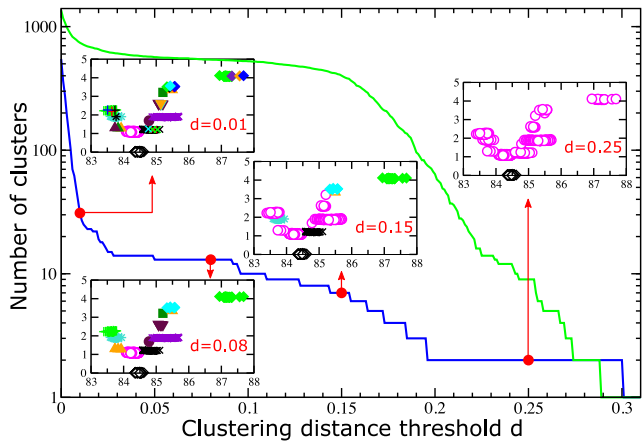}
\caption{ (color) \textbf{The number of clusters as a function of the distance threshold  $\rm{d}$} for all structures (green line)
and low-energy structures within approximately 4 kJ/mol of the ground state (blue line) for the case of $Z=3$.
Insets show the enthalpy (kJ/mol) as a function of volume ($\AA^3$) per molecule for different values of $\rm{d} = 0.01,\ 0.08,\ 0.15,\ 0.25$.
Different colors and point types in each inset correspond to different clusters.
The value of $\rm{d} = 0.05 - 0.09$ can distinguish different clusters successfully.
For $Z=2$ and $Z=4$, see Supplementary Material Fig.S1}
\label{fig3}
\end{figure}

In Fig.\ref{fig3} we display a step by step clustering analysis
where a bottom-up distance-based hierarchical clustering approach
with single linkage is used to identify the unique polymorphs among
all the structures obtained with CSP. In distance-based approaches, a
similarity metric is defined so that a distance can be measured between
data points, and clusters are constructed based on proximity. In this
study we use as the metric the fingerprint-based cosine distance \cite{USPEX-molecule,USPEX-fgpr}
defined in the EA code USPEX \cite{USPEX}:
\begin{equation}
 D_{\mathrm{cosine}}(1,2) = \frac{1}{2}\left(1-\frac{F_1 \cdot F_2}{|F_1|\;|F_2|}\right),
\end{equation}
where individual structure fingerprints are defined as
\begin{equation}
F_{AB}(R) = \sum_{A_i,\mathrm{cell}} \sum_{B_j} \frac{\delta(R-R_{ij})}{4\pi R_{ij}^2 \frac{N_A N_B}{V} \Delta} - 1,
\end{equation}
where the double sum runs over all $i$th molecules of type $A$ within the unit cell and
all $j$th molecules of type $B$ within a distance $R_{\mathrm{max}}$;
$\delta(R-R_{ij})$ is a Gaussian-smeared delta function;
$R_{ij}$ is the distance measured from the centers of molecules $i$ and $j$;
$V$ is the unit cell volume;
the function $F_{AB}(R)$ is discretized over bins of width $\Delta$;
$N_A$ and $N_B$ are the number of molecules of type $A$ and $B$, respectively.

The distance threshold used to define whether two
data points belong to the same cluster is then monotonically increased.
As a result the cluster population evolves from the situation where
every data point forms a distinct cluster to the situation in which all
data points belong to the same global cluster, revealing the bottom-up
and hierarchical nature of the approach. Translated to the CSP problem,
this data mining approach transforms the challenge of identification of
unique polymorphs from the visual comparison of all structures into an easier
decision on the value of the distance-threshold. The optimal distance threshold is such that each data cluster
matches a unique physical polymorph. In the case of Glycine a
distance threshold around 0.05-0.1 is found to be appropriate to identify the
low energy polymorphs successfully (see Supplementary Material Table.S1 and related .cif files).
The so-determined optimal threshold can
serve in advanced supervised learning techniques and be fed back in the
CSP procedure to increase considerably the efficiency by reducing the generation of replicas of already
explored structures.

\begin{figure}[!h]
    \centering
\includegraphics[width=0.47\textwidth]{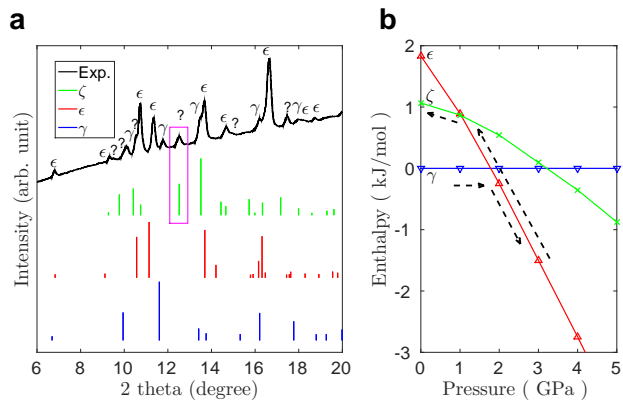}
\centering
\caption{ (color) \textbf{Assignment of the $\zeta$ phase.}
\textbf{a}. Comparison of simulated x-ray diffraction patterns for $\epsilon$-, $\gamma$- and $\zeta$ - glycine
at 2 GPa with experimental data taken from \cite{zeta} at 0.2 GPa.
The XRD of proposed $\zeta$-glycine can explain most of the unassigned peaks
that were marked in the experimental spectrum.
The theoretical spectra are calculated at higher pressure to offset the overestimation of ground state volumes in ab initio calculations.
\textbf{b}. Enthalpy per molecule as a function of pressure for $\epsilon$-glycine and $\zeta$-glycine with respect to the $\gamma$ phase up to 5 GPa.
The black arrows indicate the phase transitions observed in the experiment \cite{zeta}: Under pressure, the $\gamma$ phase undergoes a phase transition to $\epsilon$-glycine.
The decompression of $\epsilon$-glycine instead results in the $\zeta$ phase.
}
\label{fig4}
\end{figure}

The cluster analysis outlined above identifies all experimentally
observed phases of glycine compatible with 2, 3 or 4 molecules per cell,
as well as suggesting others, hereon named according to their enthalpy-per-molecule ordering.
Phases 1 to 11 lie within approximately 2~kJ/mol of the experimentally most stable phase, $\gamma$.
Among them one of the lowest energy polymorphs (phase 2) can be
identified with $\zeta$--glycine based on the excellent agreement with
XRD results (Fig.\ref{fig4}(a)) as well as its pressure evolution (Fig.\ref{fig4}(b)).
The structural identification of the $\zeta$ phase, previously experimentally observed but
not resolved up to now, marks an important achievement for CSP and is a key result of our study.
The search for $\alpha$--glycine proved very demanding despite
it being the experimentally most readily formed polymorph at ambient conditions.
In this study the $\alpha$ phase could not be found even after 20 generations
with the standard settings in USPEX.
This difficulty revealed one of the remaining challenges of CSP: the effective exploration of the topology of
an erratic and vast configuration space. Indeed, the fully ab-initio scheme advocated for in this work
pays for the higher accuracy with a heavy computational cost that makes this effectiveness even more crucial.
To improve on this aspect we weighted the random selection of the space group
of the candidate structures according to the frequency distribution appearing in known organic crystal structure database
[$P2_1/c$ (36.59~\%), $P\overline{1}$ (16.92~\%), $P2_12_12_1$ (11.00~\%), $C2/c$ (6.95~\%), $P2_1$ (6.35~\%), $Pbca$ (4.24~\%),
and uniform otherwise] \cite{Baur1992}.
This procedure successfully produced the $\alpha$ phase at the $14^{\mathrm{th}}$
generation, demonstrating that incorporation of even mild and system
unspecific experimental knowledge in the search strategy may have a
significant impact to overcome the effectiveness challenge in the most demanding cases.

Indeed if more system specific information is available it can be used to further
constrain and guide the phase-space search: limiting the search to the
experimentally known, $P2_{1}/c$, space group of $\alpha$--glycine, or
fixing the cell shape to its experimental value, resulted in its identification
at the $15^{\mathrm{th}}$ and $8^{\mathrm{th}}$ generations, respectively. Combining the
two constraints resulted in an even quicker discovery at the third
generation.

Once the low energy structures are found and examined, the configuration
space search can be further instructed to look for certain patterns. In
the case of $\alpha$--glycine, it is noteworthy that the crystal building
block can be seen as a glycine dimer, with head to tail orientation.
This feature is not seen in other ambient pressure polymorphs of glycine,
and can be speculated to be one of the reasons for the $\alpha$ phase
not being readily connected with other phases in the energy landscape.
This correlates with the difficulty of generating the structure during
the EA procedure,
as well as with its exceptional stability under pressure. Instead, if
the dimer unit is taken as building block in a CSP search, the $\alpha$
phase is found at the third iteration and new phases such as phase 8, phase 14, phase 24 and
phase 38 are also discovered.

Hence the difficulty of exploring the $\alpha$ phase as well as the
finding of new phases only after a dimer unit is employed, underlines the
remaining challenges of CSP and calls for even more efficient methods
for exploring new structures and innovative data analysis applications to
guide the search on the go for a full optimization of resources.
(see Supplementary Material Fig.S2-10 for details of all search attempts).

\section{Conclusion}
We presented a fully blind, fully ab initio crystal
structure prediction test on Glycine, a system that has been examined
several times in the past yet never fully grasped. A remarkable precision
and a broad sampling is obtained in an affordable computational time
thanks to last generation van der Waals density functionals and
evolutionary algorithms at the leading edge. The comparison of our
results with existing experimental studies enabled us to resolve
the so-far unidentified $\zeta$ phase a decade after its first
experimental observation. Further analysis of the results of the blind
test allowed us to propose several new thermodynamically plausible structures
with varying volume, compressibility and polarization. To address the
experimentally well established but CSP-wise challenging $\alpha$ phase,
we introduced an intuitive sampling strategy based on crystal structure
relative frequency found in nature. This strategy successfully found this challenging phase
and allowed us further insight in the energy landscape. Overall, the results of our blind test
shows us that a reliable crystal structure prediction procedure is possible with
incorporation of several complementary recipes to reach success, emphasizing
that one-size-fits-all solutions are yet to be discovered. Fortunately,
the leap in precision and sampling capability we have demonstrated with
these new generation tools opens new paths for crystal structure
prediction with data processing procedures such as clustering algorithms.
Hence we strongly believe ab initio CSP as presented here has come a long way
and that a new standard for structure prediction
for molecular crystals is set, and an interdisciplinary horizon for
computational science within this field is now open.

\subsection*{Acknowledgments}
Work supported by the Italian MIUR through the PRIN 2010 initiative (PRIN 20105ZZTSE).
Computational resources have been provided by SISSA and CINECA, Italy, and on Curie@TGCC-CEA
through PRACE Project 2011050736.


\end{document}